\documentclass[reprint,superscriptaddress,showkeys,amsmath,amssymb,prl,longbibliography,citeautoscript]{revtex4-2}

\usepackage[dvipsnames]{xcolor}

\usepackage{amsmath}
\usepackage{amssymb}
\usepackage[hidelinks]{hyperref}
\usepackage{chemformula}
\usepackage{diffcoeff}
\usepackage{siunitx}
\usepackage[capitalise]{cleveref}

\usepackage[normalem]{ulem}
\usepackage{color,xcolor}

\DeclareSIUnit{\angstrom}{\textup{\AA}}
\newcommand{\vct}[1]{\boldsymbol{#1}}

\setcitestyle{super}

\begin{document}

\title{Delta Machine Learning for Predicting\\Dielectric Properties and Raman Spectra}

\author{Manuel Grumet}
\affiliation{Physics Department, TUM School of Natural Sciences, Technical University of Munich, 85748 Garching, Germany}
\author{Clara von Scarpatetti}
\affiliation{Physics Department, TUM School of Natural Sciences, Technical University of Munich, 85748 Garching, Germany}
\author{Tomáš Bučko}
\email{tomas.bucko@uniba.sk}
\affiliation{Department of Physical and Theoretical Chemistry, Faculty of Natural Sciences, Comenius University
in Bratislava, SK-84215 Bratislava, Slovakia}
\affiliation{Institute of Inorganic Chemistry, Slovak Academy of Sciences, SK-84236 Bratislava, Slovakia}
\author{David A. Egger}
\email{david.egger@tum.de}
\affiliation{Physics Department, TUM School of Natural Sciences, Technical University of Munich, 85748 Garching, Germany}

\begin{abstract}
\section*{Abstract}\vskip -10pt
\noindent Raman spectroscopy is an important characterization tool with diverse applications in many areas of research.
We propose a machine learning method for predicting polarizabilities with the goal of providing Raman spectra from molecular dynamics trajectories at reduced computational cost. 
A linear-response model is used as a first step and symmetry-adapted machine learning is employed for the higher-order contributions as a second step. 
We investigate the performance of the approach for several systems including molecules and extended solids. 
The method can reduce training set sizes required for accurate dielectric properties and Raman spectra in comparison to a single-step machine learning approach.\\
\begin{center}
\textbf{TOC GRAPHIC}\vskip 12pt
\includegraphics[width=0.25\columnwidth]{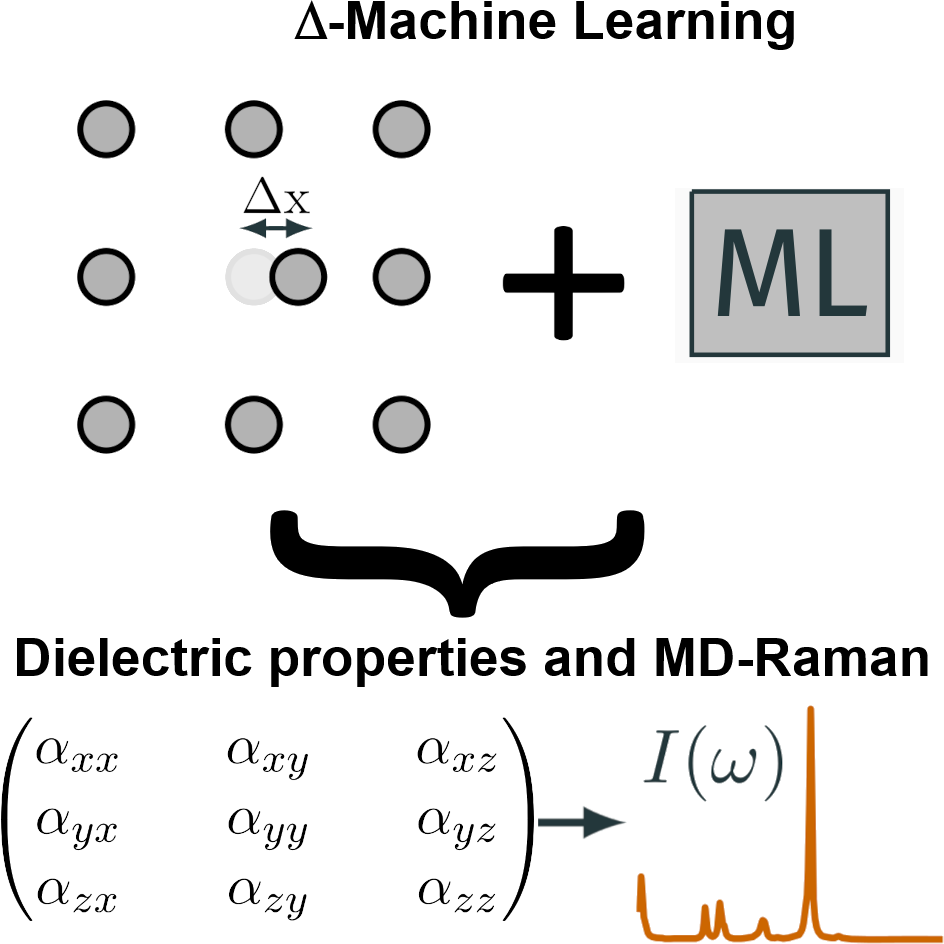}
\end{center}
\end{abstract}

\date{\today}

\maketitle

\begin{figure}
    \includegraphics[width=1.0\columnwidth]{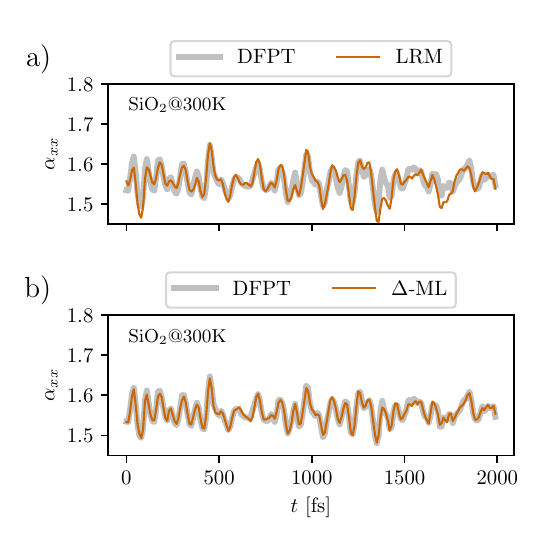}
    \caption{(a) LRM predictions of the $\alpha_{xx}$ component of the
      polarizability tensor in \ch{SiO2} over a period of \SI{2000}{fs},
      compared to DFPT reference data. (b) $\Delta$-ML predictions for the
      same data, obtained with $N_t = 50$ and $N_v = 10$.
      Note that the values shown here are polarizability per volume.}
    \label{fig:timeseries}
\end{figure}

\begin{figure}
    \includegraphics[width=1.\columnwidth]{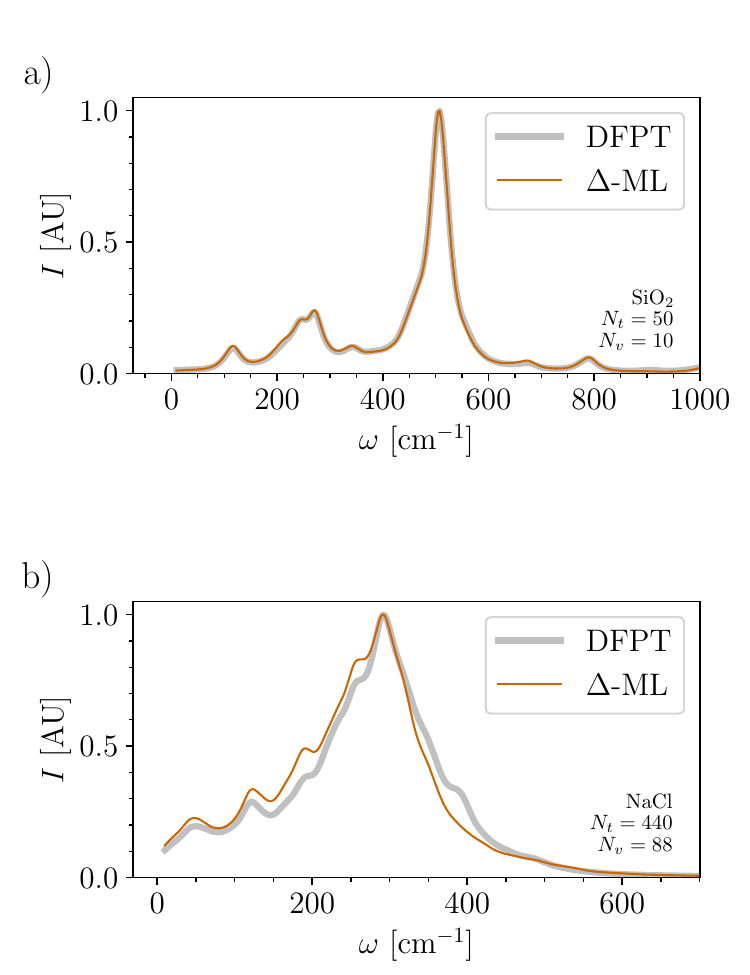}
    \caption{(a) Raman spectrum for \ch{SiO2} computed from $\Delta$-ML
      predictions, compared to DFPT reference spectrum.
      (b) Raman spectrum for \ch{NaCl} computed from $\Delta$-ML predictions, compared
      to DFPT reference spectrum.}
    \label{fig:spectrum}
\end{figure}

\noindent 
Atomic motions are often key to physical and chemical phenomena occurring at finite temperature, both in solid-state and molecular systems.
Experimentally, 
the dynamical behavior of such systems can be probed by Raman
spectroscopy. It is a table-top technique that is available in many 
laboratories because it is less complicated and expensive than, for example, neutron 
scattering~\cite{jones_etal_2019}.
Raman spectroscopy plays an important role in many areas of research, including catalysis \cite{hartman_etal_2016, loridant_2021, hess_2021, yoo_etal_2023}, perovskites photovoltaics
\cite{yaffe_etal_2017, ibaceta-jana_etal_2020, huang_etal_2022, cohen_etal_2022} and semiconductor physics more broadly \cite{yu_cardona_fundamentals_of_semiconductors},
and ionic conductors \cite{julien_etal_2018, famprikis_etal_2021, brenner_etal_2022}.
Computational predictions of Raman 
spectra using first-principles calculations are an important counterpart to experimental measurements. 
They provide further insight into the behavior of materials,  
facilitate the interpretation of measured spectra \textit{via}
theory-experiment comparisons, and enable predictions of dynamical 
properties in new compounds.\\
The central quantity for theoretical calculations of Raman spectra is the
polarizability tensor, $\vct{\alpha}$. It describes the first-order
dielectric response of a system to external electric fields.
In practice, the dielectric tensor, $\vct{\epsilon}$, can be used for periodic systems since it contains the same information~\cite{angyan_vdw}; in the following text, we shall use both quantities interchangeably. 
Raman spectra are commonly calculated within the harmonic
approximation whereby the derivatives 
of $\vct{\alpha}$ with respect to atomic displacements, determining the intensity of peaks, are calculated along 
eigenvectors of harmonic modes~\cite{togo_tanaka_2015, skelton_etal_2017, popov_etal_2020}. 
But this method is limited since it cannot capture
anharmonic effects, higher-order Raman scattering, or the explicit
temperature dependencies of Raman modes. These effects are relevant 
in a variety of physical systems and scenarios. For example, a description of phase-transitions 
in solid materials requires temperature-dependent phonon modes~\cite{dove_lattice_dynamics}, which cannot be captured in a strictly harmonic phonon picture.\\
Molecular dynamics (MD) simulations offer a way to include these effects and overcome limitations of the harmonic approach. A Raman spectrum
can be computed from an MD trajectory by calculating Fourier-transformed
velocity autocorrelation functions of the components of $\vct{\alpha}$~\cite{putrino_parrinello_2002, thomas_etal_2013, ditler_luber_2022}. However, this requires
computing a time series of $\vct{\alpha}$ values from multiple 
MD snapshots. The number of data points needed in
such an MD-Raman approach depends on the desired frequency resolution and the
total range of frequencies that needs to be covered, but typically at
least a few hundred points are needed. Such polarizability calculations can be done from first-principles
using density functional perturbation theory (DFPT)~\cite{baroni_resta_1986}, but this renders MD-based Raman calculations
computationally demanding.
The large computational efforts involved in MD-Raman calculations limit the range of physical scenarios and systems one can investigate with the method.\\
The computational cost of first-principles
calculations can be significantly reduced by machine learning (ML) techniques. 
Specifically, ML can exploit redundancies and symmetries of structures generated by MD and is now
widely used for learning and predicting energies and forces~\cite{noe_etal_2020, jinnouchi_etal_2019, unke_etal_2021}.
More recently, these methods have also been applied to predict tensorial
properties including polarizabilities, which currently is a very active research area~\cite{han_etal_2022}.
Both symmetry-adapted kernel-based methods
\cite{raimbault_etal_2019,
wilkins_etal_2019,
lewis_etal_2023,
fang_etal_2023,
inoue_etal_2023,
litman_etal_2023}
and neural-network approaches
\cite{
sommers_etal_2020,
zhang_etal_2020,
tuan-anh_zalesny_2020,
shang_wang_2021,
nguyen_lunghi_2022,
zhang_jiang_2023}
have been used for this,
as well as a physically-based small parametric model~\cite{paul_etal_2023}.
In addition, ML methods have also been applied to compute aspects of
Raman spectra directly, without explicit consideration of polarizabilities
\cite{gandolfi_etal_2020, ren_etal_2021, petrusevich_etal_2023}.
Delta ML
($\Delta$-ML) is a combined approach to predicting physical quantities:
a computationally inexpensive approximation is used as a first
step and ML methods are then applied to learn only the \textit{differences
between first-step predictions and true values}~\cite{ramakrishnan_etal_2015}. 
The challenge in any $\Delta$-ML method 
is the search for a physical model to achieve sound first-step predictions that 
can seamlessly be combined with the proceeding ML method.
It has been previously shown that the prediction of polarizabilities in a molecular crystal can be improved by using the polarizability of the molecular monomer as a first step \cite{raimbault_etal_2019}.
However, to the best of our knowledge, no systematic $\Delta$-ML approach that can be applied beyond the case of molecular crystals has been proposed and investigated yet for prediction of dielectric properties such as polarizabilities and Raman spectra.
\\
In this letter, we propose a $\Delta$-ML method for predicting dielectric properties of molecules and materials. 
Focusing on polarizabilities, we suggest a linear response model (LRM) that encodes key information on the dielectric response of the system as a first step.
Combination of the LRM with ML for tensorial properties is assessed \textit{via} inspection of polarizability components and Raman spectra.
We find that the $\Delta$-ML method increases accuracy and reduces required training-set size compared to a direct ML approach where the same ML model is applied to the polarizability data directly, i.e., without using any first-step approximation.
Applying $\Delta$-ML to small molecules and extended solids including more complicated materials, as well as discussing its inherent limitations and potential for further improvements, we demonstrate its predictive power for practical MD-Raman calculations across a broad range of physical systems.\\
In order to develop our method, we start from a Taylor expansion of a component of $\vct{\alpha}$ with respect to
atomic displacements ($\Delta \vct{x}$) from their respective equilibrium positions or any other reference structure ($\vct{x}_{0}$):
\begin{equation}
  \label{eq:taylor}
  \alpha_{\alpha\beta}(\vct{x}) = \alpha_{\alpha\beta}(\vct{x}_0) + \sum_{i=1}^{3N_\text{at}} \diffp{\alpha_{\alpha\beta}}{x_{i}} \Delta x_{i} + \cdots
\end{equation}
where $N_\text{at}$ is the number of atoms of the system and the 
index $i$ enumerates the components of a $3N_\text{at}$ dimensional atomic position vector $\vct{x}$. 
\cref{eq:taylor} is an exact formula which can be approximated to arbitrary order.
Here, we consider the simplest, first-order variant: 
a LRM is constructed by determining the constant term, $\vct{\alpha}(\vct{x}_{0})$, and the
respective derivatives, $\partial \alpha_{\alpha \beta} / \partial x_{i}$. The constant
term can be determined \textit{via} a single DFPT calculation for the
equilibrium structure. 
The first-order derivatives can be obtained through additional DFPT
calculations on displaced structures.
Applying a central difference
method, two calculations are necessary
for each atomic degree of freedom.
We note that the problem of choosing the coordinate frame in systems with rotational degrees of freedom can easily be solved by employing rotationally and translationally invariant internal coordinates, as discussed in Sec.~I.E of the Supporting Information (SI). 
A numerical demonstration that rigid rotations do not significantly deteriorate predictions of our LRM model is given in Sec. I.H of the SI.
Symmetry
considerations can be used in this procedure in order to reduce the number of required calculations, since
any two symmetry-equivalent atoms imply derivatives that are related \textit{via} similar symmetry operations as position vectors of atoms.
Thus, a total of $6N+1$ calculations are required to parameterize the
LRM, where $N$ is the number of symmetry-inequivalent atoms in the
system. 
Further details about the LRM are presented in the SI.\\
We parameterize the LRM \textit{via} DFPT calculations~\cite{gajdos_etal_2006} using \texttt{VASP}~\cite{kresse_furthmuller_1996} and the PBE exchange-correlation functional~\cite{perdew_etal_1996}.
This allows for predicting polarizabilities of MD snapshots by extracting the displacements $\Delta
x_{ij}$ from it and applying \cref{eq:taylor}.
\cref{fig:timeseries}a shows the performance of
predictions with the LRM for the $\alpha_{xx}$
component in \ch{SiO2} along a selected portion of an MD trajectory at 300\,K, which we obtain using density functional theory (DFT) in \texttt{VASP}.
Compared to the DFPT reference data for the same MD trajectory, the LRM accurately captures many of the $\alpha_{xx}$ oscillations.\\
\begin{figure}
    \includegraphics[width=1.0\columnwidth]{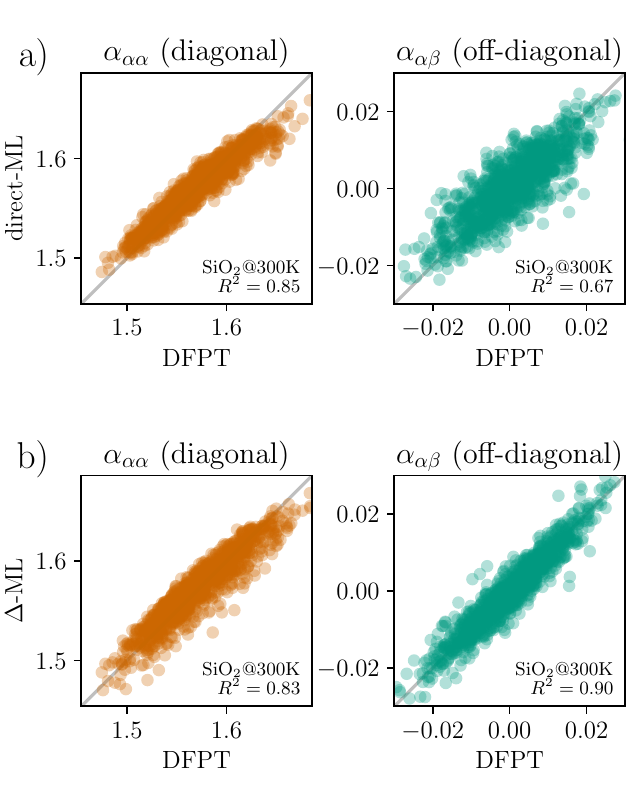}
    \caption{(a) Scatterplot comparing direct-ML predictions of
      diagonal (left) and off-diagonal (right) polarizability components in \ch{SiO2} to DFPT reference data,
      obtained with $N_t = 20$ and $N_v = 4$.
      (b) Scatterplot comparing
      $\Delta$-ML predictions to DFPT, obtained with the same training
      and prediction set.
      Note that the values shown here are polarizability per volume.
    }
    \label{fig:scatterplots}
\end{figure}%
The LRM has several limitations:
first, the polarizability can have a local extremum at the reference positions, in which case the derivatives, $\partial \alpha_{\alpha\beta} / \partial
{x}_{i}$, vanish. 
This occurs when atomic motion along the respective axis is
first-order Raman inactive. 
Additionally, it is also possible that these
derivatives are zero only for certain components of $\vct{\alpha}$, which will be discussed below. In these
cases, the LRM can only capture the constant term and the predicted
polarizability timeseries for these components will therefore be
approximately constant. 
These are important limitations of the present LRM, which, in principle, can be overcome in a straightforward way by 
accounting for further terms in the expansion of \cref{eq:taylor}.
Here, we focus on the fact that these limitations motivate an ML model as an additional second step
that can capture higher-order displacement responses.\\
For the second step in our $\Delta$-ML approach we employ kernel-based methods
\cite{karniadakis_etal_2021, deringer_etal_2021}. The underlying idea
is to use a descriptor that captures relevant aspects of atomic
configurations and a kernel function that allows to quantify
similarities between different configurations along MD trajectories. These similarities can then be used
to perform fitting using kernel ridge regression (KRR). For many
physical quantities, descriptors based on overlap integrals are an
appropriate choice. A common method which takes into account
symmetries is smooth overlap of atomic positions (SOAP)
\cite{bartok_etal_2010, bartok_etal_2013, willatt_etal_2019,
musil_etal_2021}. For fitting tensorial quantities such as $\vct{\alpha}$, the extended $\lambda$-SOAP approach is well-suited because
it also takes into account tensorial covariance properties
\cite{grisafi_etal_2018}. While the choice for $\lambda$-SOAP implies no difference for some
tensor invariants such as the mean polarizability (see Sec. IC in the SI), it does provide a benefit for
fitting the off-diagonal elements of $\vct{\alpha}$.\\ 
\begin{figure*}
        \includegraphics[width=1.0\columnwidth]{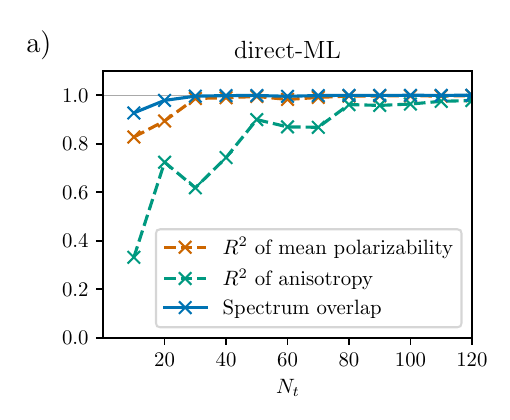}
    \includegraphics[width=1.0\columnwidth]{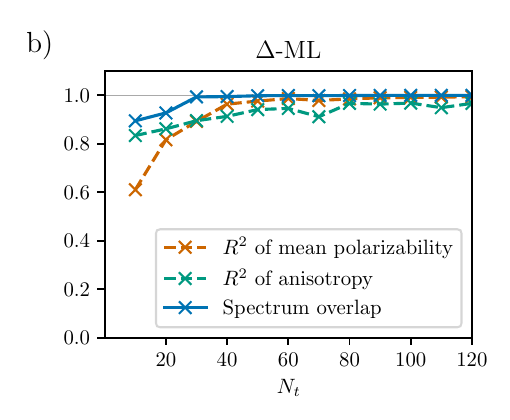}
    \caption{(a) Performance metrics for direct ML predictions in
      \ch{SiO2} as function of training set size $N_t$. (b) The same
      performance metrics for $\Delta$-ML predictions.
      }
    \label{fig:scores}
\end{figure*}%
We perform the fitting procedure in KRR with $\lambda$-SOAP using the \texttt{dscribe} \cite{himanen_etal_2020} and \texttt{librascal} \cite{musil_etal_2021a} packages together with
\texttt{scikit-learn} \cite{pedregosa_etal_2011}.
The training data for the ML model are obtained from DFPT
calculations on a subset of MD trajectories. Note that the $6N+1$ 
configurations that were required for the LRM can, in principle, be reused as training data for the ML model, which we did not attempt here to ease the comparison. Furthermore, a
validation dataset is selected from the MD trajectory in order to optimize ML
hyperparameters with respect to validation error. For
the sizes $N_t$ and $N_v$ of the training and validation set, respectively, we used a
constant ratio $N_t:N_v = 5:1$. This ensures that the validation set is scaled
in proportion whenever the number of training data points is increased.
In the spirit of $\Delta$-ML, LRM predictions are
always subtracted from DFPT values before feeding them to ML, such that only differences are learnt.
Further details on the fitting procedure can be found in Sec. I.F of the SI.\\
\cref{fig:scatterplots} shows scatterplots of a polarizability timeseries for \ch{SiO2} predicted with $\Delta$-ML and an otherwise identical ML approach without the LRM (direct-ML), comparing both to reference DFPT calculations along an MD trajectory.
The diagonal components of $\vct{\alpha}$ are predicted well already with direct-ML compared to the DFPT reference data. 
For this particular case, even LRM alone achieved fairly accurate predictions (\textit{cf.}~\cref{fig:timeseries}a), implying that the diagonal components are relatively easy to capture for \ch{SiO2}. However, the off-diagonal components are predicted far less accurately in direct-ML. 
Remarkably, we find that our $\Delta$-ML method provides similarly accurate predictions for both the diagonal and off-diagonal components of $\vct{\alpha}$ (see \cref{fig:scatterplots}b and \cref{fig:timeseries}b).
The findings suggests that the $\Delta$-ML approach allows for more efficient learning of all components of $\vct{\alpha}$  because the LRM provides a good first approximation for dynamic fluctuations of this quantity.\\
To assess the accuracy of our method, we compared it again to direct-ML for the case of \ch{SiO2}, using DFPT results as a reference and calculating the coefficient of determination, $R^2$ (see SI for details). 
We focus on the tensor invariants $a$ (mean polarizability) and $\gamma^2$ (anisotropy) of $\alpha$ (see Sec. IC of the SI), since these are directly relevant for Raman calculations, and show $R^2$ for \ch{SiO2} as a function of $N_t$ in \cref{fig:scores}.
We find that for achieving $R^2$ close to 1 for both $a$ and $\gamma^2$, $\Delta$-ML requires $N_t$ to be on the order of only 20, which outperforms direct-ML by at least a factor of 2.
Thus, the LRM and ML methods encoded in $\Delta$-ML complement each other well and may offer accurate predictions of $\alpha$ with smaller training-set sizes.\\
The size of the training set required to obtain good prediction performance is expected to strongly depend on the system.
Therefore, we investigate the versatility of $\Delta$-ML by computing $R^2$ for a broader range of physical systems that include extended solids and gas-phase molecules.
The $N_t$ values that are required to achieve good prediction performance are listed in \cref{tab:scores} for each system.\\
The solid AlN is an interesting, challenging example because of its known LO/TO splitting that makes prediction of its dielectric and Raman properties a difficult problem~\cite{popov_etal_2020}.
Thus, AlN illustrates several complications that require an accurate computational methodology for prediction of its Raman spectrum.
For this case as well as for \ch{SiO2}, we find that $\Delta$-ML achieves a significant reduction of required $N_t$ compared to direct-ML.
The small gas-phase molecules (\ch{H2O}, \ch{CH4}, and \ch{CH3OH}) we consider here are found to require similarly low $N_t\sim10$ in both approaches, further demonstrating the broad applicability of the $\Delta$-ML method.\\
We consider two more extended systems, Si and NaCl, in order to demonstrate how the aforementioned inherent limitations of the chosen LRM are compensated by the proceeding ML step: 
the change of the diagonal components of $\vct{\alpha}$ with $\Delta x_i$ is approximately an even function in Si (see SI), for which the simple LRM predicts a constant timeseries for the $\alpha_{\alpha\alpha}(\vct{x}(t))$ values that does not capture the pertinent temporal fluctuations in the system.
NaCl is another challenging case because it is not first-order Raman active and the LRM therefore merely predicts $\alpha_{\alpha\beta}(\vct{x}(t))=\textrm{const}$.
\cref{tab:scores} shows that  $\Delta$-ML and direct-ML lie on par for these two cases, \textit{i.e.}, the proceeding ML can compensate for the lack of dynamical information in the underlying LRM. 
Hence, even when, for physical reasons,  the choice of our specific LRM may not provide any benefit for learning components of $\vct{\alpha}$, it does not worsen prediction performance.
Therefore, the $\Delta$-ML approach can seamlessly integrate a simple physical model and ML procedure in order to capture the temporal and spatial fluctuations of $\vct{\alpha}$.
These findings suggest that $\Delta$-ML is a promising approach for dielectric predictions of dynamical systems. 
Furthermore, since the LRM is the simplest approximation to \cref{eq:taylor}, it can still be extended in a straightforward way to further improve prediction performance of $\Delta$-ML if needed.\\ 
\begin{table}[b]
  \caption{Minimum required training set sizes to achieve $R^2 > 0.8$ in both $a$ (mean polarizability) and $\gamma^2$ (anisotropy) of $\alpha$ for different systems. $N_t/N_v$ was kept constant at 5:1, $N_t$ was increased in steps of 20 for AlN and NaCl and in steps of 10 for all other systems, and $R^2$ was evaluated on separate data sets of size 400.}
  \begin{ruledtabular}
  \begin{tabular}{lcc}
               & Direct ML      & $\Delta$-ML  \\\hline
\ch{SiO2}      &             50 &             20 \\
AlN            &            360 &            280 \\
Si             &             60 &             60 \\
NaCl           &            440 &            440 \\\hline
\ch{H2O}       &             10 &             10 \\
\ch{CH4}       &             20 &             10 \\
\ch{CH3OH}     &             10 &             10 \\
  \end{tabular}
  \end{ruledtabular}
  \label{tab:scores}
\end{table}%
Close connections of the here discussed dielectric quantities to Raman spectra are established \textit{via} a correlation-function analysis. 
Specifically, calculation of the
Raman spectrum requires the Fourier-transformed velocity autocorrelation
functions (VACFs) of the tensor components of $\vct{\alpha}$, \textit{i.e.},
\begin{equation}
\langle \dot\alpha_{\alpha\beta}(\tau) \, \dot\alpha_{\alpha\beta}(\tau+t) \rangle_\tau = 
  \int_{-\infty}^{\infty} \dot\alpha_{\alpha\beta}(\tau) \, \dot\alpha_{\alpha\beta}(\tau+t) \dl \tau.
  \label{eq:vacf}
\end{equation}
The terms $a_{\tau}$ and $\gamma_{\tau}^2$  can then be computed from the
VACFs~\cite{thomas_etal_2013, long_the_raman_effect}, see Sec. ID in the SI for details. A spherically-averaged Raman spectrum can then be obtained as
\begin{equation}
  I(\omega) \propto
  \frac{(\omega_\text{in} - \omega)^4}{\omega}
  \frac{1}{1 - \exp\left(-\frac{\hbar \omega}{k_B T}\right)}
  \frac{45a_{\tau}^2 + 7\gamma_{\tau}^2}{45}.
  \label{eq:intensity}
\end{equation}
Note that the frequency-dependent prefactor in \cref{eq:intensity} is
not exact. Several other versions of this equation also exist,
which can be derived based on different approximations for taking
into account the quantum nature of the atomic motion~\cite{egorov_skinner_1998}. \\
We applied our $\Delta$-ML method to calculate MD-Raman spectra at
\SI{300}{K} for the test systems described above, using full DFPT calculations as a reference. 
\cref{fig:spectrum} showcases two examples and the other systems are discussed in the SI.
While our focus is on the performance of the $\Delta$-ML method, we point out that the spectra are also affected by other aspects of the computational setup, such as the choice of density functional approximation.
Good agreement between $\Delta$-ML and DFPT is obtained for \ch{SiO2} as expected from our above findings and further quantified through 
the calculated cosine similarity, $S_C$, shown as spectrum overlap in \cref{fig:scores} (see SI).
The case of NaCl is particularly interesting since, as we noted above, it is a first-order
Raman inactive material, which means that only
higher-order effects contribute to the Raman spectrum.
For this reason, the spectrum contains contributions from $q$-points other than $\Gamma$ \cite{benshalom_etal_2022}.
This makes MD-Raman calculations particularly challenging, requiring a
large simulation cell in order to provide sufficient sampling of $q$-space (see SI for details). 
In addition, we find that the ML procedure required relatively large $N_t$ in this case. We 
speculate that $N_t$ could be reduced \textit{via} improvement of the first step in the $\Delta$-ML procedure, for example by including higher-order terms. 
With sufficient accuracy of the terms $a$ and $\gamma$, the Raman spectrum for NaCl agrees well with DFPT and recent experimental data~\cite{benshalom_etal_2022}. To our knowledge, this is
the first ML-based MD-Raman calculation for a higher-order Raman material.\\
It is interesting to note that for MD-Raman calculations we observed a certain insensitivity on errors in
the individual components of $\vct{\alpha}$. 
Specifically, relatively inaccurate predictions of tensor components can still result in a relatively accurate Raman spectrum (see spectrum overlap in \cref{fig:scores}).
In addition, errors at different points along $\vct{\alpha}(\vct{x}(t))$ are, to a good approximation, random
and independent of each other. Thus, these errors tend to average out when computing autocorrelation functions.\\
Finally, we mention that our method provides room for further adaptation in future work. 
Possibilities include use of more advanced first-step models instead of the LRM, different descriptors in the ML step, and changes to the hyperparameter optimization scheme such as employing cross validation.\\
In conclusion, we proposed a $\Delta$-ML approach that unifies a
physical model with symmetry-adapted ML for prediction of dielectric 
properties and Raman spectra and is applicable to a broad range of different
systems. We focused on the polarizability tensor and chose a simple LRM as a starting point to describe the dynamic dielectric fluctuations, which is completed by an ML procedure in a second step. 
The $\Delta$-ML method can perform better than an 
otherwise identical direct-ML approach with the same training set size. This is especially because the
LRM step provides a benefit for predicting off-diagonal components of the polarizability tensor. Since
the data points needed for parameterizing the LRM can be reused as ML
training data, $\Delta$-ML does not necessarily increase computational costs compared to direct-ML. 
We also investigated specific systems for which the LRM method provides no benefit by design and found that it does not deteriorate ML prediction performance for these cases.
Our findings show that $\Delta$-ML is a promising approach for predictions of dielectric properties and Raman spectra of molecules and materials at finite temperature.
It provides a way to reliably compute spectra that capture the full extent of atomic motions in molecules and materials without relying on the harmonic approximation at a reasonable computational cost.
We speculate that the $\Delta$-ML approach might also be useful for the calculation of other properties that require time-correlation functions, such as infrared spectra or transport coefficients. \\

\begin{acknowledgements}
\noindent C.v.S. thanks Felix Schwarzfischer for hepful discussions.
Funding provided by the Alexander von Humboldt-Foundation in the
framework of the Sofja Kovalevskaja Award, endowed by the German Federal
Ministry of Education and Research, by TUM.solar in the context of the Bavarian
Collaborative Research Project Solar Technologies Go Hybrid
(SolTech), and by TU Munich - IAS,
funded by the German Excellence Initiative and the European Union
Seventh Framework Programme under Grant Agreement No. 291763, are
gratefully acknowledged.
The authors further acknowledge the Gauss Centre for Supercomputing e.V.
for funding this project by providing computing time through the John
von Neumann Institute for Computing on the GCS Supercomputer JUWELS at
Jülich Supercomputing Centre.
Part of the research was obtained using the computational resources procured in the national project National competence centre for high performance computing within the Operational programme Integrated infrastructure (project code: 311070AKF2).
T.B. acknowledges support from Slovak Research and Development Agency under the Contract No. APVV-20-0127.
\end{acknowledgements}
\vskip -20pt~\section*{Supporting Information Statement}

\noindent Further computational details and additional results on displacement dependence of the polarizability in silicon, fitting performance and training set size as well as additional Raman spectra.

\vskip -20pt~\section*{Notes}

\noindent Code availability: The programs needed to perform the calculations presented in this work can be obtained upon reasonable request from the corresponding authors.

\vskip -20pt
\section*{References}\vskip -20pt
\bibliographystyle{achemso}
\bibliography{references.bib}

\end{document}